\documentclass{article}
\usepackage{amsfonts,amssymb,amsmath,graphicx}

\begin{document}

\title{The Geometry and Topology of Data and Information for Analytics of 
Processes and Behaviours: Building on Bourdieu and Addressing New Societal 
Challenges}

\author{Fionn Murtagh \\
fmurtagh@acm.org}

\maketitle

\begin{abstract}
We begin by summarizing the relevance and importance of inductive analytics 
based on the geometry and topology of data and information. Contemporary 
issues are then discussed.  These include how sampling data for representativity 
is increasingly to be questioned.  While we can always avail of analytics 
from a ``bag of tools and techniques'', in the application of machine
learning and predictive analytics, nonetheless we present the case for Bourdieu
and Benz\'ecri-based science of data, as follows.  This is to construct bridges 
between data sources and position-taking, and decision-making. There is summary 
presentation of a few case studies, illustrating and exemplifying application domains.  
\end{abstract}

\section{Introduction}

From the data analytics viewpoint, a summary of the main procedural principles and 
features of the analytical methodology arising out of Pierre Bourdieu's work
is provided in Lebaron (2015).  The following points are related to that summary.
\begin{enumerate}
\item Determine and display social field structure, or even more comprehensively, 
social space configuration.
\item Show structural homologies between fields or social spaces, based on 
axis, or factor, interpretation. 
\item So, this is reasoning by analogy, therefore
inductive reasoning, but in a context that can use, in an integral manner, 
statistical modelling and predictive learning. The latter encompasses machine 
learning methods.
\item Determine relative autonomy between fields or social spaces.  This 
is through use of comparative procedures, and causal hypotheses on the 
relations between fields. 
\item Study sub-spaces in a global social space. This is through use of 
class specific analysis, or MCA, Multiple Correspondence Analysis of 
sub-clouds of subjects.
\item Informed by statistics, and supervised (with training set, test set) machine 
learning, explain social practice and position-taking,
with associated decision-making.  
\item Effects and consequences, thus predictive field-related outcomes.
Here, there may be integrated use of analysis of variance (ANOVA) or regression, or
of supervised machine learning methods. 
\item Study field dynamics using hierarchical classification, in factor space, 
use of appropriate supplementary or contextual elements (attributes, subjects)
that can form structuring factors, and fully integrate time-evolving data. 
\end{enumerate}

This comprehensive and far reaching data analysis methodology, will encompass 
statistical modelling, machine learning, predictive analytics, and visualization 
analytics, whenever these are relevant and appropriate.  Relative to current machine 
learning practice that can be characterized as the ``bag of tools and techniques'' 
approach to analytics, the following are unquestionably major benefits.  Field 
and homology, and 
underpinning these, geometry and topology, employed for inductive reasoning, and 
relating such reasoning with position-taking or decision making.  


Themes of this article encompass the following: application of analytics drawing 
on social fields and homology, and beyond; general and broad relevance and applicability;
underpinning the analytics, we may consider that we have: data leading to information, 
leading to knowledge, leading to wisdom. Geometric data analysis (GDA), and based on 
GDA, topology, is the essential basis for all such work.

In section \ref{sect2}, we both justify and strongly motivate Bourdieu-related 
analytics for contemporary data analytics problems.  

In section \ref{sect3}, in our analytics, we are seeking explanatory or elucidatory
narratives.  Three case studies are at issue.  Firstly, there is the expression 
or manifestation of underlying emotions.  Secondly, there is the convergence of 
events or activities, expressing causal or consequential outcomes.  Thirdly, there 
are the most influential patterns and trends in very large data sources.

Section \ref{sect4} introduces particular issues in, and perspectives on, the analytics 
of behaviour and activity.  This is pursued with a preliminary study in section \ref{sect5}.

\section{New Challenges and Opportunities in the Context of Big Data Analytics}
\label{sect2}

The following expands on our contribution to the discussion in Keiding and Louis (2016).
The comprehensive survey (with 141 references) of Keiding and Louis (2016) sets out 
new contemporary issues of sampling and population distribution estimation.  An 
important take-home message
is this: ``There is the potential for big data to evaluate or calibrate survey 
findings ... to help to validate cohort studies''.  Examples are discussed of ``how 
data ... tracks well with the official'', and contextual, repository or holdings.  It is 
well pointed out how one case study discussed ``shows the value of using `big data' 
to conduct research on surveys (as distinct from survey research)''.  This arises
from what has been discussed by Japec et al., (2015): ``The new paradigm means it is now 
possible to digitally capture, semantically reconcile, aggregate, and correlate data.''

Limitations though are clear: 
``Although randomization in some form is very beneficial, it is by no means a 
panacea. Trial participants are commonly very different from the external ... 
pool, in part because of self-selection, ...''.  This is due to, ``One type of 
selection bias is self-selection (which is our focus)''.  

Important points towards addressing these contemporary issues include the following.
``When informing policy, inference to identified reference populations is key''.
This is part of the bridge which is needed, between data analytics technology and 
deployment of outcomes.  
``In all situations, modelling is needed to accommodate non-response, dropouts and 
other forms of missing data.''

While ``Representativity should be avoided'', here is an essential way to address
in a fundamental way, what we need to address:  ``Assessment of external validity, 
i.e.\ generalization to the population from which the study subjects originated or 
to other populations, will in principle proceed via formulation of abstract
laws of nature similar to physical laws''.

Interesting perspectives that support Keiding and Louis (2016), include
Friedman et al.\ (2015), and the following, from Laurison and Friedman (2015): 
``... the GBCS [Great British Class Survey] data have three important 
limitations. First, the GBCS was a self-selecting web-based survey, ... This means 
it is not possible to make formal inferences. ...  the nationally representative 
nature of the Labour Force Survey (LFS) along with its detailed and accurate measures 
... facilitates a much more in-depth investigation ...''  In a blog posting, 
Laurison (2015) points very clearly to how, just ``Because the GBCS is not a 
random-sample or representative survey'', other ways can and are being found to 
draw great benefit.  

Another different study on open, free text questionnaires (Z\"ull 
and Scholz, 2011, see also 2015) notes selection bias, but also: 
``However, the reasonable use of 
data always depends on the focus of analyses.  So, if the bias is taken into 
account, then group-specific analyses of open-ended questions data seem appropriate''. 

The bridge between the data that is analyzed, and the calibrating Big Data, is 
well addressed by the geometry and topology of data.  Those form the link between
sampled data and the greater cosmos.  Bourdieu's concept of field is a prime
exemplar.  Consider, as noted by Lebaron (2009), how Bourdieu's work, involves
``putting his thinking in mathematical terms'', and that it ``led him to a conscious
and systematic move toward a geometric frame-model''.  This is a multidimensional,
``structural vision''.  Bourdieu's analytics ``amounted to the global [hence Big Data]
effects of a complex structure of interrelationships, which is not reducible to the 
combination of the multiple [... effects] of independent variables''.  The concept
of field, here, uses Geometric Data Analysis that is core to the integrated data 
and methodology approach used in the Correspondence Analysis platform (used, in 
general, in Murtagh, 2010; and comprehensively in Le Roux, 2014).

An approach to drawing benefit from Big Data is precisely as described 
in Keiding and Louis (2016).  The noting of the need for the ``formulation of abstract 
laws'' that bridge sampled data and calibrating Big Data can be addressed, for the 
data analyst and for the application specialist, as geometric and topological.  

The principles and practice following from the Big Data setting for our analytics
can also be linked to ethical issues.  Consider, for example, that sources of data 
start with: interviews, online media, and big data including data that is unstructured, 
structured, heterogeneous multimedia, open linked, and so on.  One example is to 
consider the long term prediction of criminal activity, versus the prediction of 
exposure or vulnerability to crime.  The latter can lead to preemptive and 
preventative action, with no ethical aspects involved.  A narrative approach, 
therefore using analytical field dynamics, offers a practical and effective 
alternative (Harcourt, 2002, 2006).  Over and above the sole use of quantitative
prediction, we are noting the importance of integrated qualitative and quantitative 
analytics.  

In line with Harcourt's work (Harcourt, 2002), with the central role of Correspondence
Analysis, this programme of work encompasses: not ``content analysis'' but rather 
``map analysis''.  I.e.\ ``cognitive mapping, relational analysis, and meaning 
analysis'' among other names.   Harcourt notes the need ``to visually represent the 
relationship between structures of social meaning and the contexts and practices 
within which they are embedded''.

Harcourt's work has case studies related to sexual activity, and activities with 
firearms. We would propose the great importance of pursuing field and homology analytics, 
of underpinning geometric and topological data analysis, for all levels of digital,
online or cyber crime and misdemeanour.  
From Harcourt (2005): ``Where do we stake the boundary of the criminal law -- 
or, more importantly, how? How do we decide what to punish? Do we distribute these 
vices, these recreations, these conducts -- what do we even call these things? --
into two categories, the passable and the penal, and then carve some limiting 
principle to distinguish the two? Are we, in the very process, merely concocting 
some permeable line -- a Maginot line -- to police the criminal frontier?''

In this section, we have noted the contemporary very wide-ranging, and comprehensive, 
applicability of analytics that come from the work of Pierre Bourdieu, taking the 
integral theory and practice of Jean-Paul Benz\'ecri's life work.  Note has also 
been made of how this can incorporate and include the mostly sole use of ``bag of 
tools and techniques'' (informally expressed in this way) characterization of 
unsupervised and supervised machine learning.  


\section{Narrative Analytics, and Narrative Synthesis}
\label{sect3}

We consider a complex web of relationships.
Semantics include web of relationships -- thematic structures and patterns. 
Structures and interrelationships evolve in time.   
Semantics include time evolution of structures and patterns, including both: 
threads and commonality; and change, the exceptional, the anomalous. 
Narrative can suggest a causal or emotional relationship between events.
A story is an expression of causality or connection.
Narrative connects facts or views or other units of information.

In Murtagh et al.\ (2009), there is stochastic analysis of structure and style, 
of the film script of the movie, Casablanca, shot by Warner Brothers between May and
August 1942.  This both statistically and qualitatively supports McKee's (1999) 
statement that the composition of Casablanca is ``virtually perfect''.  

An application that resulted from 
this work was visualization of narrative, supporting collective, collaborative 
narrative construction.  This related to collective book authoring by literature 
students in university, and for many books published, by school children, following 
this narrative display and support framework.  See Reddington et al.\ (2013).  

\begin{figure}
\includegraphics[width=9cm]{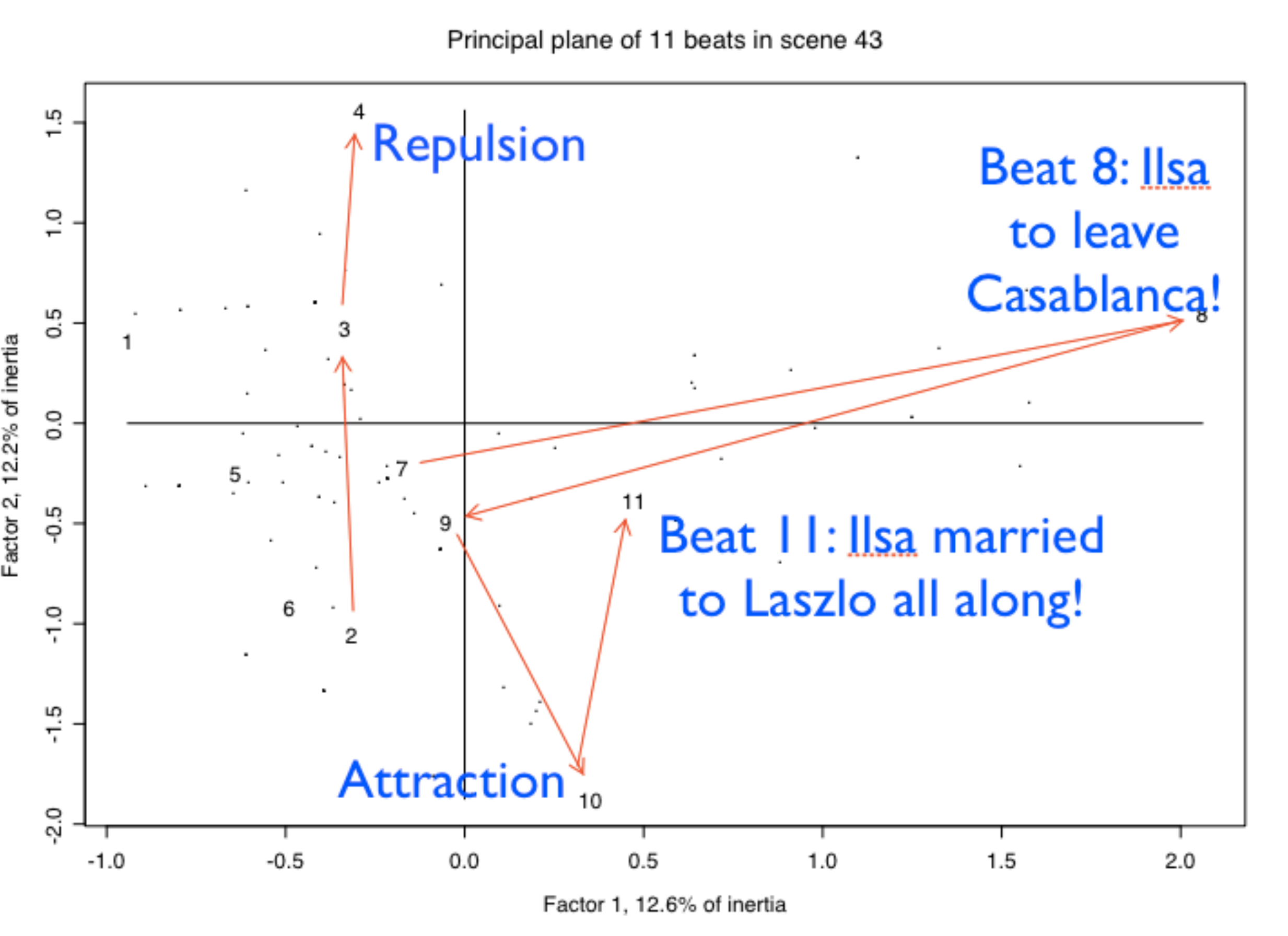}
\caption{Correspondence Analysis of the scene 43, which crosses (with
presence/absence values) 11 successive beats (numbers here) with, in total,
210 words (dots: not labelled for clarity here). Two of the beats are 
annotated, as is axis 2.}
\label{fig1}
\end{figure}

Figure \ref{fig1} displays and depicts McKee's description of ``mid-act climax'',
scene 43, subdivided into 11 ``beats'' or subscenes.  This portrays emotional 
attraction and repulsion between Rick and Ilsa, on the second axis (respectively, 
here, negative, positive directions); the first axis counterposes externality 
relative to the main location, Casablanca.  

In the tracking of emotion, we are determining and tracking emotion in 
an unsupervised way.  (This is not machine learning like in sentiment analysis, which 
is supervised.) Emotion is understood as a manifestation of the unconscious.  
Social activity causes emotion to be expressed or manifested.  (Cf. Murtagh, 2014.)

We used chapters 9, 10, 11, 12 of Gustave Flaubert's 19th century novel, Madame Bovary.  
This concerns the three-way relationship between Emma Bovary, her husband Charles, 
and her lover Rodolphe Boulanger.

\begin{figure}
\includegraphics[width=9cm]{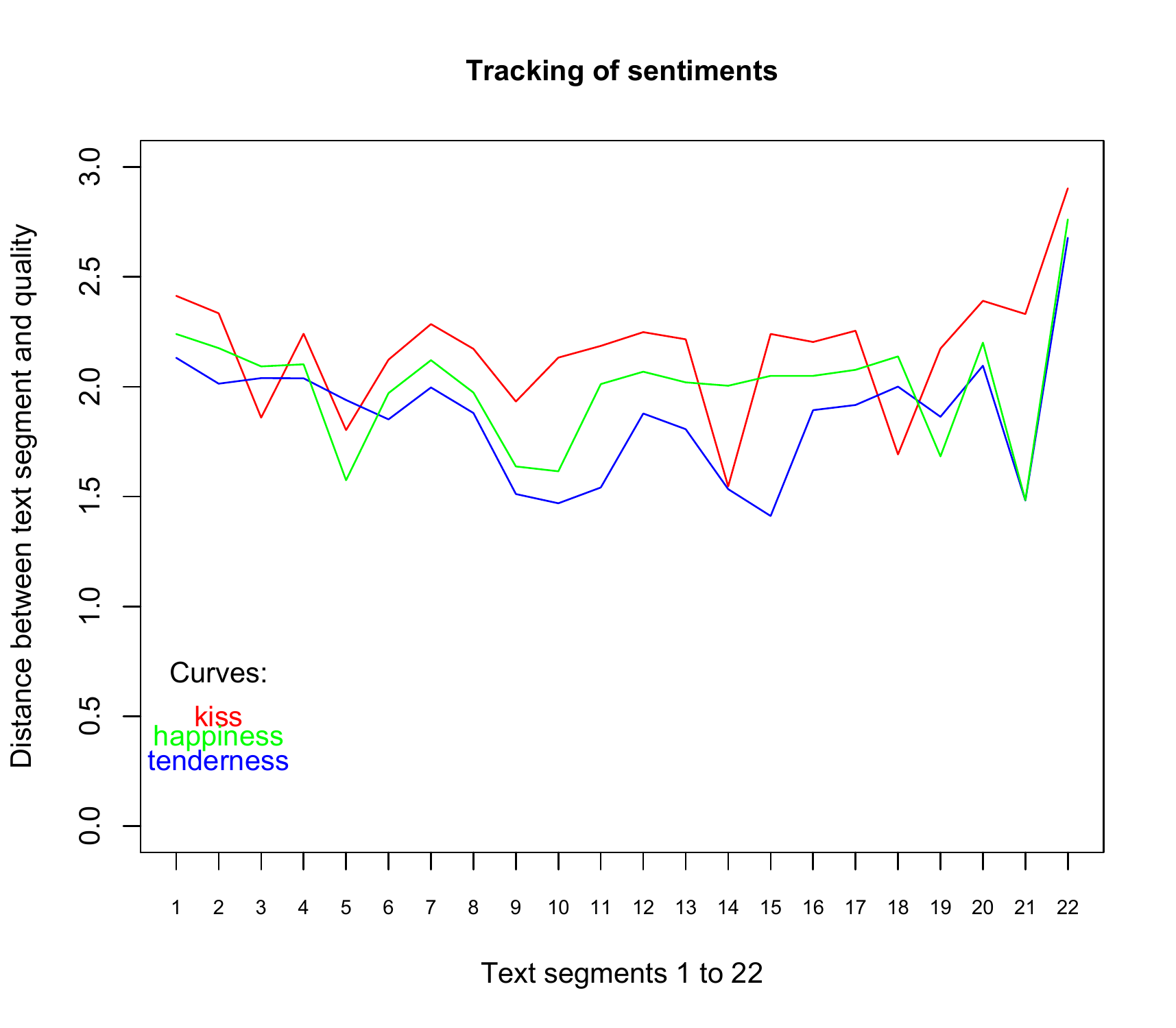}
\caption{A low value of the emotion,
 expressed by the words ``kiss'', ``happiness'' and
``tenderness'', implies small distance to the text segment.
The chronology of sentiment tracks the closeness of these different
sentimental terms relative to the narrative, represented by the text segment.  Terms
and text segments are vectors in the semantic, factorial space, and the full
dimensionality of this space is used.}
\label{fig2}
\end{figure}

In Figure \ref{fig2}, there is display of the evolution of sentiment, expressed 
by (or proxied by) the terms ``kiss'', ``tenderness'', and ``happiness''.   We see 
that some text segments are more expressive of emotion than are other text segments.

In Murtagh and Ganz (2015) we look at, firstly,
emotional interaction in the Casablanca movie,
using dialogue (and dialogue only) between main characters Ilsa and Rick.
Secondly, we look at all of the text in chapters 9 to 12 of the novel, 
Madame Bovary.  In both studies, we used the same methodology, i.e.\ cross-tabulation 
of word sets,
comprising the entire universe of discourse, and including function and grammar words
that characterized textual ``texture''.  We find the latter to be very useful for
expression of emotion. 

Figure \ref{fig2} displays the evolution of sentiment, expressed by (or
proxied by) the terms ``kiss'', ``tenderness'', and ``happiness''.   We see
that some text segments are more expressive of emotion than are other text
segments.

We now briefly describe another study of both quantitative and qualitative 
analysis.  

A particular narrative pattern is at issue now, arising out of a social media case study, 
and motivated by J\"urgen Habermas's {\em Theory of Communicative Action} 
({\em Theorie des kommunikativen Handelns}).  
For impact of actions, here based on initiating tweets on Twitter, related to 
environmental issues, this is considered as: semantic distance between the initiating 
action, and the net aggregate outcome.  This can be statistically modelled and tested.

\begin{figure}[htb]
\includegraphics[width=9cm]{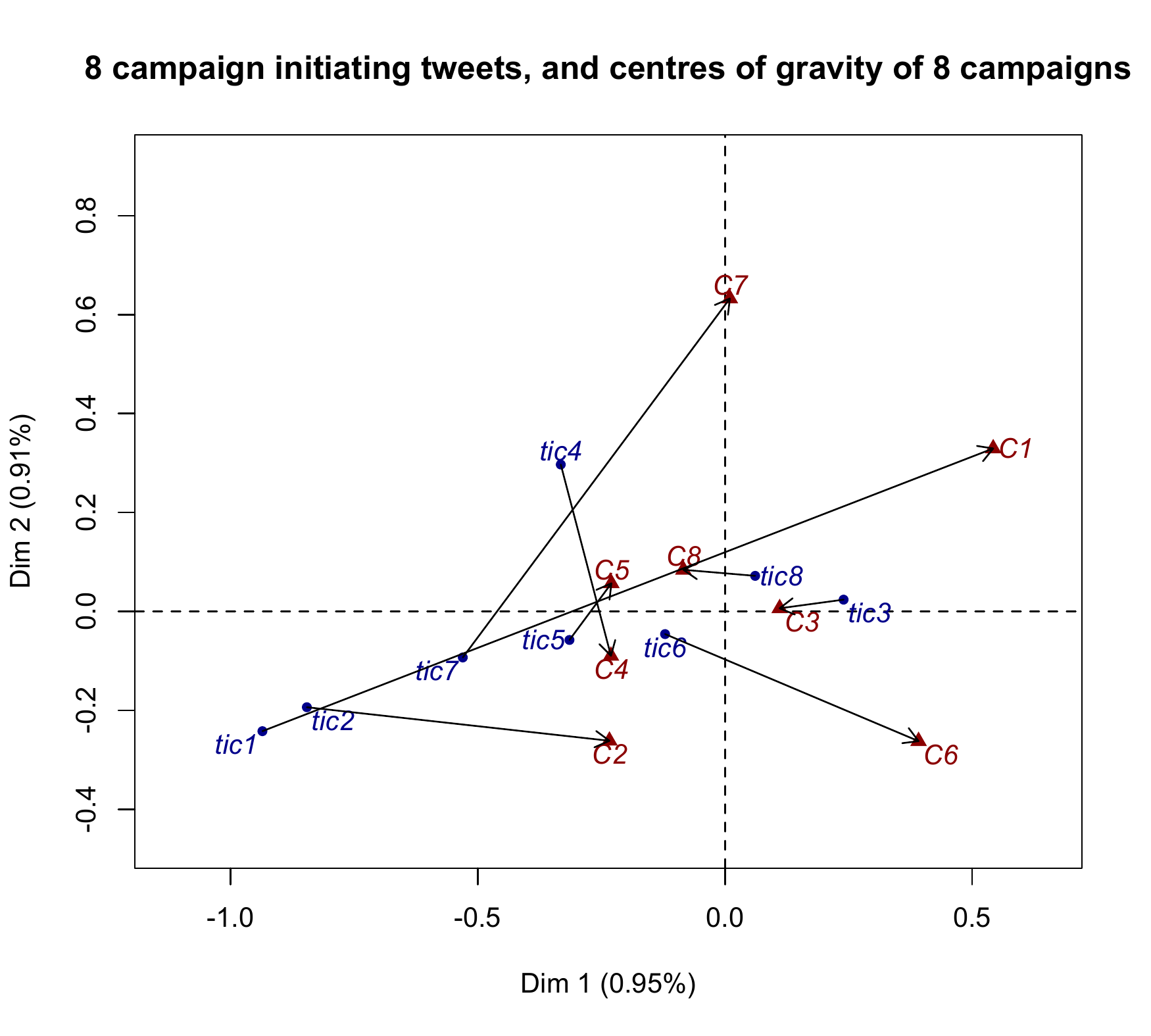}
\caption{The campaign initiating tweets are labelled ``tic1'' to ``tic8''.
The centres of gravity of the campaigns, i.e.\ the net aggregate of the campaigns, are
labelled ``C1'' to ``C8''.  In each case, the tweet initiating the campaign is
linked with an arrow to the net aggregate of the campaign. The percentage inertia
explained by the factors, ``Dim 1'' being factor 1, and ``Dim 2'' being factor 2,
is noted. This was also studied in the full, factor space, dimensionality.}
\label{fig3}
\end{figure}

Figure \ref{fig3} (Murtagh et al., 2016), shows 
the 8 tweets that initiated the campaigns, and the net aggregate
campaigns, given by the centres of gravity of the 8 campaigns.
We see that campaigns 3, 5, 8 have initiating tweets that are fairly close to the
net overall campaign in these cases.
By looking at all tweets, and all terms, it is seen that the campaign
initiating tweets, and the overall campaign means, are close to the origin, i.e.\
the global average.  That just means that they (respectively, initiating tweets,
and means) are relatively unexceptional among all tweets.
While the information that we find in our data is very faint, nonetheless we have an
excellent visualization of this information, that we further analyze, 
quantitatively. 

A further study is at issue, next, with a far larger Twitter data source.  

\begin{figure}[htb]
\includegraphics[width=10cm]{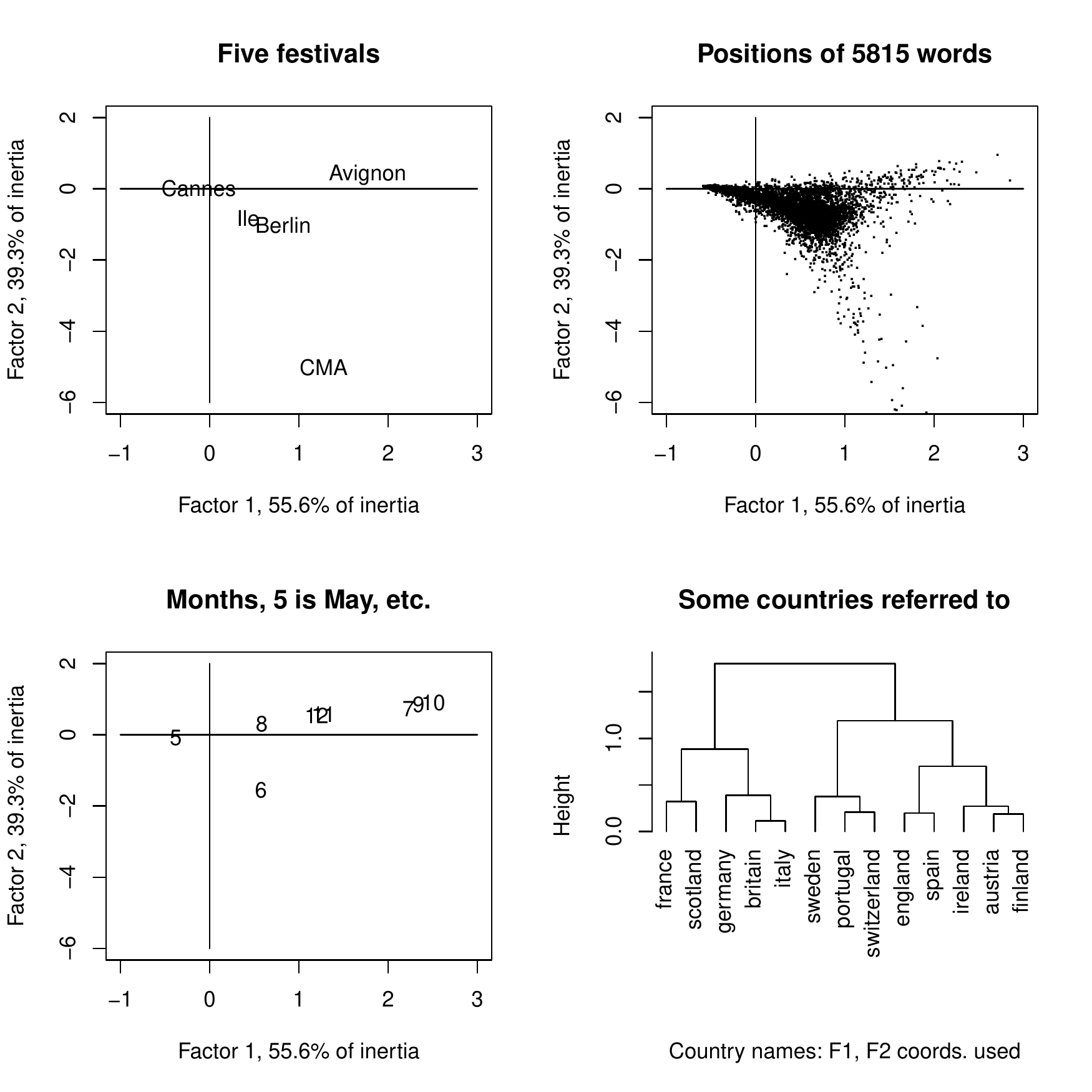}
\caption{From Twitter data in May to December 2015, 
relating to (film, music, etc.) festivals.  The principal factor plane, 
accounts for just 95\% of the inertia or information content.  The hierarchical
clustering of the country names is based on their locations, that are in the upper
left panel plot.} 
\label{fig4}
\end{figure}

In Murtagh (2016), we are concerned with social and community aspects of cultural 
festivals.  Our data is about 12 million Twitter tweets, in the time period, May to
December 2015.  (Further data will be analysed by us in the future.) Included are such 
events as the following:
Cannes Film Festival (13--24 May 2015); F\`eis \`Ile, Isle of Islay (Scotland) Festival 
(23--31 May 2015); Berlin Film Festival (19--21 May
2015); CMA, Country Music Association (Nov.\ 2015); Yulin Dog (June 2015);
and Avignon Theatre Festival (4--25 July 2015).
In our initial exploratory analyses, the critical Twitter debate in regard to the
Yulin Dog Meat Festival meant that Twitter data was quite distinct.  So in Figure 
\ref{fig4}, there is a selection of the other festivals, listed above.  A sufficiently
used word set was derived from the Twitter tweet set.  For the data used here, 
a 5815 word corpus was used, from the 32,507,477 words in the data.  Occurrence
of 1000 uses of a word was required.  Filtering was also carried out of prepositions,
parts of verbs, just some abbreviations or partial words, and non-Roman scripts 
including Cyrillic, Chinese, Japanese, and Greek.  In what was retained, 
English, French and Spanish dominated.  

An initial analytics decision to be made is what is the principal set of 
individuals or records to be used, with other such objects being supplementary 
elements; and similarly, what is the principal attribute or variable or variable
modality set to be used, with other variables being supplementary elements.  
Contextual attributes will naturally be supplementary elements.  We can have the 
following analytics perspective: we use an analytics resolution scale, that is 
such that tweets per day are considered.   Thus the principal data crosses 
the set of 233 days (from 11 May 2015 to 31 December 2015) by the retained word
corpus.  That is under investigation in Murtagh (2016).  

In Figure \ref{fig4}, the principal variables are the five festivals.  Factor 1 
is dominated by (contribution to the axis inertia) the Avignon Theatre festival, 
factor 2 is dominated by the Nashville, Tennessee,  Country Music Association, 
factor 3 is dominated by the Berlin film festival.  Word projections are shown
as dots, just indicating the semantic density.  (The numbers of occurrences of 
these festival names, which could have multiple occurrences in a single tweet, 
and certainly possibly lots of occurrences in a day's tweets, are: Cannes, 
1,615,550; Avignon, 102,499; CMA, 9407; Berlin, 5557; and \`Ile, 1998.)  The months, 
5 being May, 6 being June, and so on, until 12 being December, are displayed as
supplementary elements.  

In order to look in an exploratory way at country names used in the Twitter data, 
a set of names was selected.  This included, just to see if there were a 
distinction, England and Britain.  In the lower right panel in Figure \ref{fig4},
it is interesting how the semantics here, i.e.\ the Twitter social media sourcing, 
and also taking the principal factor coordinates, factors 1, 2, results in 
Austria and Finland, and then Ireland, being similar here.

From the point of view of analytical procedure, in the foregoing, we have made use
of the following: 
resolution scale of the analysis, and our primary focus in the analysis; 
contextual or ancillary information associated as supplementary elements; 
and selective pattern determination and trend following.  

\section{Towards Behaviour and Activity Analytics, for Mental Health, 
Depression, and Lifestyle Analytics}
\label{sect4} 

We are using Euclidean geometry for semantics of information.  In the referenced 
articles, in previous sections, we used hierarchical topology for other 
aspects of semantics, and in particular how a hierarchy expresses anomaly or change.  
A further useful case is when the hierarchy respects chronological or other sequence 
information. In our view, analytics based on Bourdieu's work, based on MCA, Multiple
Correspondence Analysis, (encompassing e.g. field, homology, habitus, etc.) 
should be a main analytics approach in many current areas of work, including smarter 
cities, analytics of Internet of Things, security and forensics (including trust and 
identity), Big Data, etc.   

It is noted in Kleinman et al.\ (2016) how relevant and important mental health 
is, given the integral association with physical health.  From Kleinman et al.\ 
(2016) there is the following: ``... parity between mental and physical health 
conditions remains a distant ideal''.  ``The global economy loses about \$1 trillion
every year in productivity due to depression and anxiety''.  ``Next steps include 
... integration of mental health into other health and development sectors''.  

In Cooper et al.\ (2016), page 4, under the heading of ``Five Ways to Wellbeing'', 
reference is made to ``mental capital and wellbeing''.  On page 14, a section is 
entitled ``The `mental capital' values of the outdoors''.

\section{Mental Health: Adult Psychiatric Morbidity Survey, England, 2007}
\label{sect5}

A periodic survey of mental health, HSCIC (2009), is used.  There are 1704 
variables, including questioning of the subjects about symptoms and disorders,
psychoses and depression characteristics, anti-social behaviours, eating 
characteristics and alcohol consumption, drug use, and socio-demographics, 
including gender, age, educational level, marital status, employment status, 
and region lived in.

As a first analysis, the following variables were selected: 14 questions, hence
14 categorical variables, relating to ``Neurotic symptoms and common mental 
disorders''.  These are described in HSCIC (2009, Appendix C).  Almost all of these
variables had as question responses, whether or not there were symptoms or disorders
in the past week, one question related to one's lifetime, and one question related
to the age of 16 onwards.  Another question set was selected, relating
to socio-demographic variables, noted in the foregoing paragraph.  In this set of 
socio-demographic variables, there were 9 variables.  

\begin{figure}[t]
\includegraphics[width=9cm]{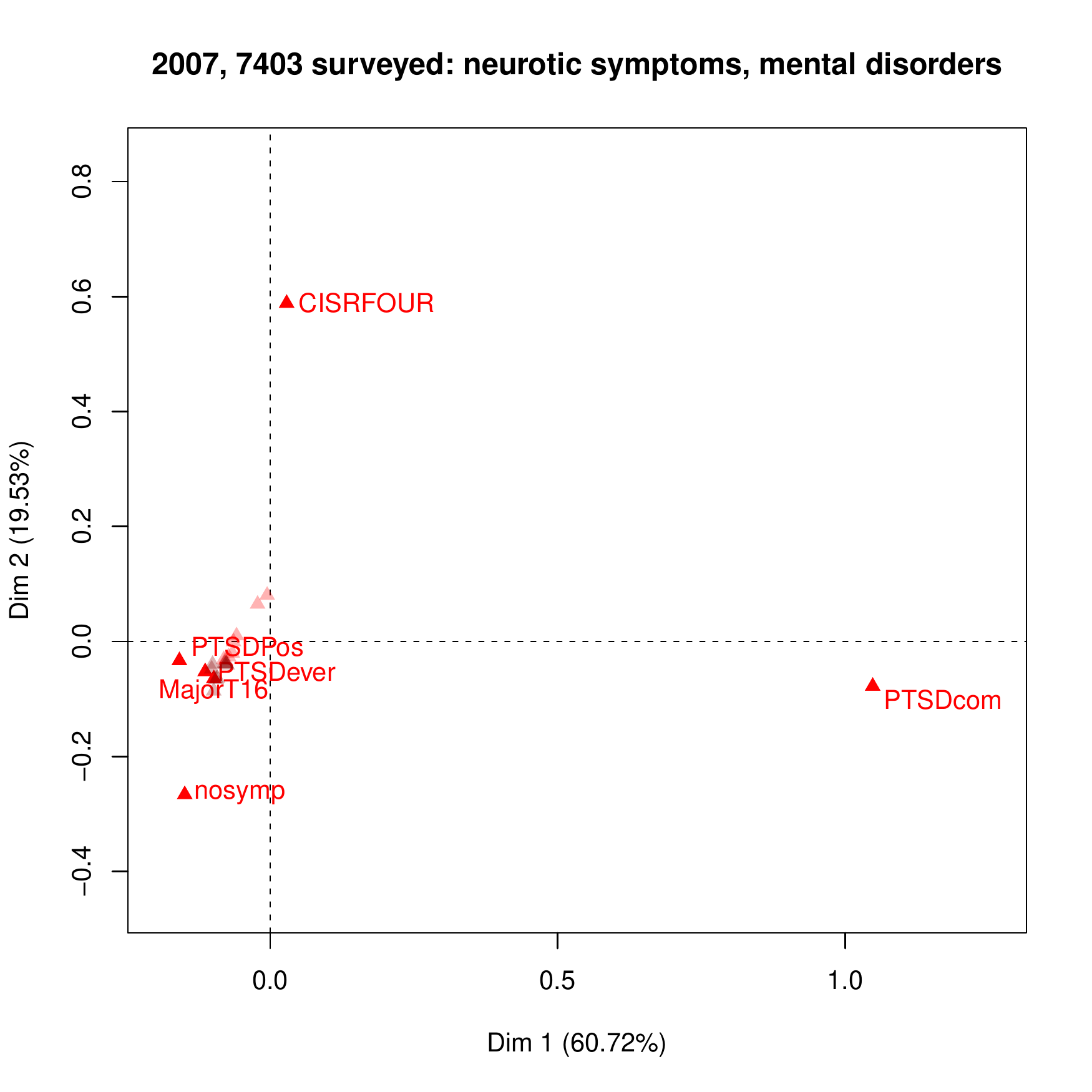}
\caption{Adult psychiatric morbidity survey 2007, England, household survey. 
The socio-demographic variables, as supplementary variables are close to the origin.
Displayed are the 6 highest contributing variables.}
\label{fig5}
\end{figure}

An initial display of the neurotic symptoms and common mental disorders, see
Figure \ref{fig5}, sought to have socio-demographic variables as supplementary.  But
these were projected close to the origin, therefore showing very little 
differentiation or explanatory relevance for the symptoms and disorders data.  

In Figure \ref{fig5}, it is found that factor 1 is explained as PTSDcom, ``Trauma 
screening questionnaire total score'' versus all other variables.  Factor 2 is 
explained as ``CISRFOUR'' versus ``nosymp''.  These are, respectively, 
``CIS-R score in four groups, 0-5, 6-11, 12-17, 18 and over. (CIS-R = Common Mental 
Disorders questionnaire)''; and no neurotic symptoms in the past week.  

\begin{figure}[htb]
\includegraphics[width=9cm]{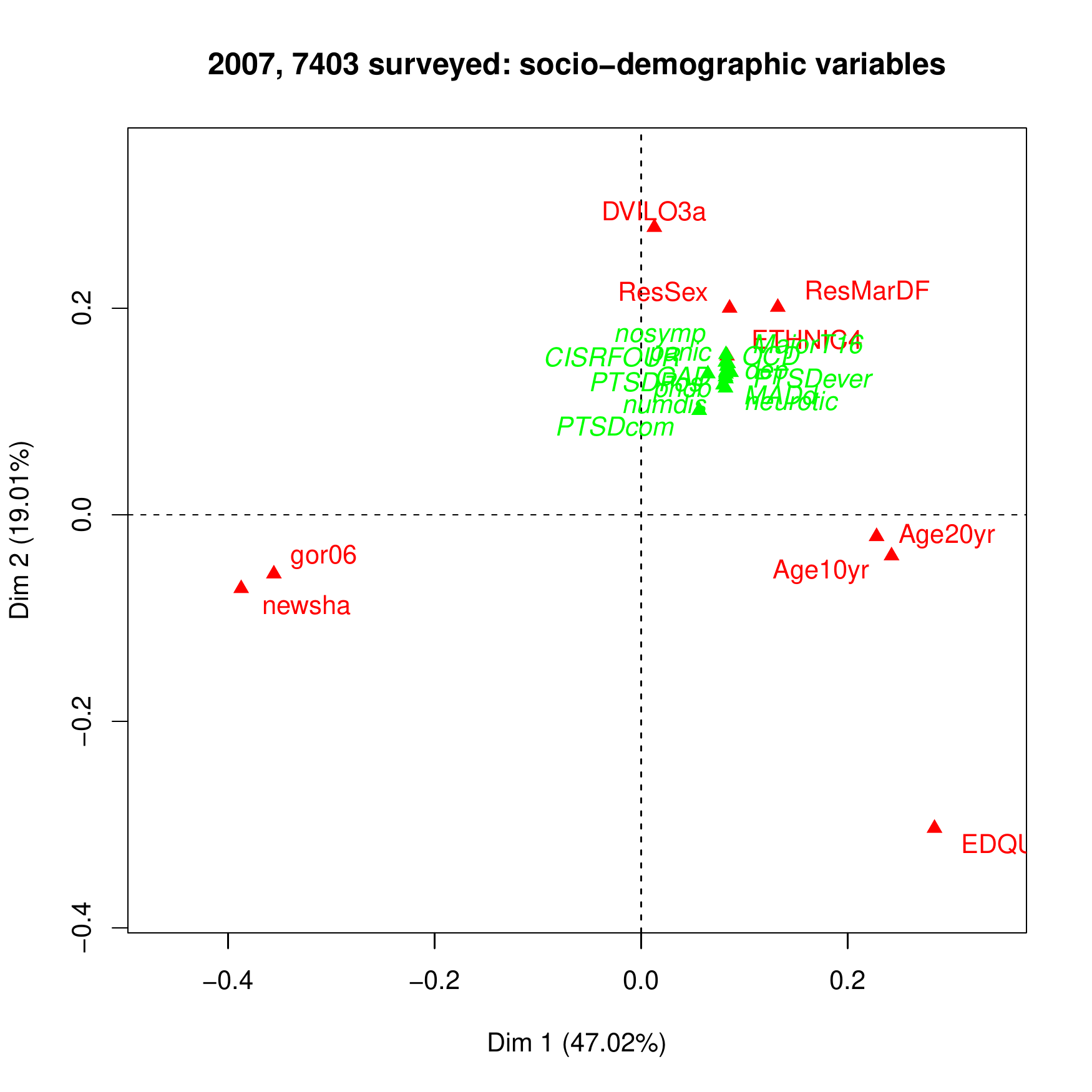}
\caption{Adult psychiatric morbidity survey 2007, England, household survey.
The neurotic symptoms and common mental disorders, as supplementary variables are
close to the origin.} 
\label{fig6}
\end{figure}

Next, Figure \ref{fig6}, it was sought to characterize the socio-demographic data, 
and then to see if the neurotic symptoms and common mental disorders data could be 
explanatory and contextual for the socio-demographic data.  But no differentiation 
was found for these supplementary variables, indicating no particular explanatory 
capability in this particular instance.  

It may be just noted how, in Figures \ref{fig5} and \ref{fig6}, the main 
and supplementary variables were interchanged.  Respectively, the symptoms 
and demographic variables were main and supplementary; then the main and supplementary 
variables were the demographic variables and symptoms.  This was done in order
to explore the data.  

In Figure \ref{fig6}, factor 1 is seen to have age and education level counterposed 
to home region.  Factor 2 is seen to have educational level counterposed to: employment 
status, gender, marital status, ethnicity.  (In the paragraph to follow, 
the names of some of these variables are explained.) 

\begin{figure}[htb]
\includegraphics[width=9cm]{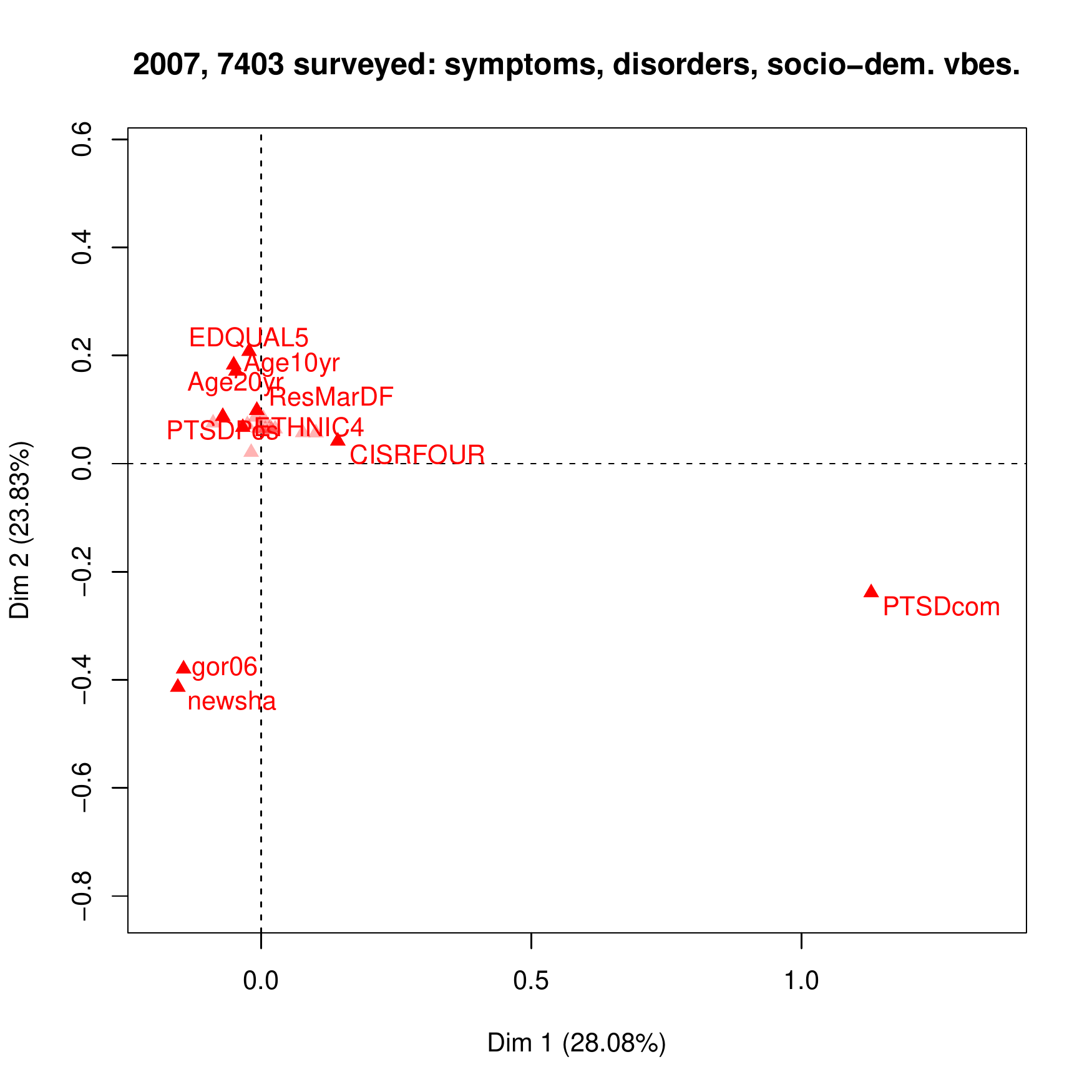}
\caption{Adult psychiatric morbidity survey 2007, England, household survey.
The analysis has the neurotic symptoms and common mental disorders, and the 
socio-demographic variables.  Displayed are the 10 highest contributing variables 
to the principal plane.}
\label{fig7}
\end{figure}

Finally, it was checked whether neurotic symptoms and common mental 
disorders data should be jointly analysed with the socio-demographic data.  Figure 
\ref{fig7} shows this outcome.  On the positive first factor, what is 
particularly important, from the contribution to the inertia, is PTSDcom, 
``TSQ (Trauma Screening Questionnaire) total score''.  The negative second factor, 
is highly influenced by these two variables: ``gor06'', ``newsha'', respectively: 
``Government office region'' and ``Strategic Health Authorities''.   These were both
sets of geographic regions in England, respectively 9 and 10.  (The 7th such region 
in each of these variables was London.)  

One aim that we have in future work is to see if we can formulate forms and 
expressions of ``mental capital''.  This concept was noted in section \ref{sect4}
where an example of it was going for a walk, therefore taking exercise,
in the countryside or parkland.  


A summary, and preliminary, interpretation derived from Figure \ref{fig7} is how
factor 1 accounts for recorded trauma, and factor 2 accounts for region of the
respondent.  

\section{Conclusions}

In this work, we have sought to provide perspectives and focus for the analytics
of behaviours and activities.  These can be related to individuals, and to social 
groupings or collectivities.  We will, in such analytics, work towards 
responding to the problem of bias in big data sampling and related representativity. 
We also seek to bridge the data with decision-making information.  In the sense of 
the latter, we are bridging data analytics with position-taking.  

Procedurally, the following can be noted.  
Methodology and implementation are quite integral to the analytics described here.  
This is counterposed to having methods used as ``black boxes'', 
which was a term used, occasionally, in the early days of neural network
methods, typically for supervised classification, or nonlinear regression.  

In this work, there is direct relationship also with psycho-analytical issues, 
with the same geometric and topological data analytics for the bi-logic 
that is manifested by the integral human reasoning and thought processes, 
both conscious and unconscious.  For background on this, and case studies, 
detailing the Correspondence Analysis platform, and ultrametric (hierarchical)
topology that is central to unconscious thought processes, see Murtagh (2014).

\section*{References}

\medskip

\noindent
Cooper C, Wilsdon J, and Shooter M. (2016)  Making the Case for the Social
Sciences, No.\ 9 Mental Wellbeing. 28 pp.  BACP, British Association for Counselling and
Psychotherapy.  Available at: \\ 
http://www.acss.org.uk/wp-content/uploads/2013/09/Making-the-Case-9-Mental-Wellbeing-Web.pdf

\medskip

\noindent
Harcourt BE (2002) Measured interpretation: introducing the method of 
Correspondence Analysis to legal studies, {\em University of Illinois Law Review}, 
979--1017. 

\medskip

\noindent
Harcourt BE (2005) Carceral imaginations, {\em Carceral Notebooks}, vol: 1. \\
Available at: http://www.thecarceral.org/imaginations.pdf

\medskip

\noindent 
Harcourt BE (2006)  {\em Against Prediction.  Profiling, Policing, and Punishing
in an Actuarial Age}. Chicago: University of Chicago Press.  

\medskip

\noindent
HSCIC, Health and Social Care Information Centre (National Health Service, UK) (2009). 
National Statistics Adult Psychiatric Morbidity in England - 2007, Results of a 
household survey, Appendices and Glossary. 174 pp. \\
Available at: http://www.hscic.gov.uk/pubs/psychiatricmorbidity07

\medskip

\noindent
Japec  L, Kreuter F, Berg M, Biemer P,  Decker P,  Lampe C,  Lane J,  
O'Neil C, and Usher A (2015) AAPOR Report on Big Data. Technical Report. 
AAPOR, American Association for Public Opinion Research, 50 pp. \\
Available at: \\
http://www.aapor.org/AAPORKentico/Education-Resources/Reports.aspx \\
and \\
https://www.aapor.org/AAPORKentico/AAPOR\_Main/media/ \\
Task-Force-Reports/BigDataTaskForceReport\_FINAL\_2\_12\_15\_b.pdf

\medskip

\noindent
Keiding N, and Louis TA (2016) Perils and potentials of self-selected entry to 
epidemiological studies and surveys. {\em Journal of the Royal Statistical Society, 
Series A} vol: 179, Part 2, 319--376.

\medskip

\noindent
Kleinman A,  Lockwood Estrin G, Usmani S, Chisholm D, Marquez PV,  
Evans TG, and Saxena S (2016) Time for mental health to come out of the shadows.
{\em The Lancet} vol: 387, 2274--2275.

\medskip

\noindent
Laurison D (2015) Blog, Three myths and facts about the Great British Class
Survey.  Available at: \\ 
http://www.thesociologicalreview.com/information/blog/ \\
three-myths-and-facts-about-the-great-british-class-survey.html

\medskip

\noindent
Laurison D, and Friedman S (2015) Introducing the class ceiling: social mobility 
and Britain's elite occupations, LSE (London School of Economics) Sociology 
Department Working Paper Series. Available at: \\
http://www.lse.ac.uk/sociology/pdf/Working-Paper\_Introducing-the-Class-Ceiling.pdf

\medskip

\noindent
Lebaron F (2009) How Bourdieu `quantified' Bourdieu: the geometric modelling of 
data.  Chapter 2 in Robson K, Sanders C (eds.), {\em Quantifying Theory: Pierre 
Bourdieu}, Heidelberg: Springer.  

\medskip

\noindent
Lebaron F (2015)  L'espace social. Statistique et analyse g\'eom\'etrique 
des donn\'ees dans l'\oe uvre de Pierre Boudieu.  (Social space.  Statistics and 
geometric data analysis in the work of Pierre Bourdieu.) Chapter 3 in Lebaron F,  
Le Roux B  (eds.), {\em La M\'ethodologie de Pierre Bourdieu en Action, Espace
Culturel, Espace Social et Analyse des Donn\'ees} (The Methodology of Pierre 
Bourdieu in Action, Cultural Space, Social Space and Data Analysis), Dunod, Paris. 

\medskip

\noindent
Le Roux B (2014) {\em Analyse G\'eom\'etrique des Donn\'ees Multidimensionnelles}, 
Paris: Dunod.

\medskip

\noindent
McKee R (1999)  {\em Story, Substance, Structure, Style, and the Principles of 
Screenwriting}, London: Methuen.  

\medskip

\noindent
Murtagh F, Ganz A, and McKie S (2009)  The structure of narrative: the case of 
film scripts.  {\em Pattern Recognition}, vol: 42, 302--312.  (Discussion in Merali Z
(June 2008) Here's looking at you, kid. Software promises to identify blockbuster 
scripts, {\em Nature}, p. 453.)

\medskip

\noindent
Murtagh F (2010) The Correspondence Analysis platform for uncovering deep 
structure in data and information, Sixth Boole Lecture, {\em Computer Journal}, vol: 53 
(3), 304--315.

\medskip

\noindent
Murtagh F (2014) Mathematical representations of Matte Blanco's bi-logic, based on 
metric space and ultrametric or hierarchical topology: towards practical application, 
{\em Language and Psychoanalysis}, vol: 3(2), 40-63.

\medskip

\noindent

\noindent
Murtagh F (2016) Semantic Mapping: Towards Contextual and Trend Analysis of 
Behaviours and Practices, in K. Balog, L. Cappellato, N. Ferro, C. MacDonald, 
Eds., {\em Working Notes of CLEF 2016 -- Conference and Labs of the Evaluation 
Forum}, \'Evora, Portugal, 5-8 September, 2016, pp. 1207-1225, 2016.
http://ceur-ws.org/Vol-1609/16091207.pdf

\medskip

\noindent
Murtagh F, and Ganz A (2015) Pattern recognition in narrative: Tracking emotional 
expression in context, {\em Journal of Data Mining and Digital Humanities}, 
vol: 2015, 21 pp. \\
Available at: http://http://arxiv.org/abs/1405.3539

\medskip

\noindent
Murtagh F, Pianosi M, and Bull R (2016)  Semantic mapping of discourse and 
activity, using Habermas's Theory of Communicative Action to analyze process, 
{\em Quality and Quantity}, vol: 50(4), 1675--1694.

\medskip

\noindent
Reddington J, Murtagh F, and Cowie D (2013)  Computational properties of fiction
writing and collaborative work, {\em Advances in Intelligent Data Analysis XII}, 
Lecture Notes in Computer Science, vol: 8207, pp.\ 369--379.

\medskip

\noindent
Z\"ull C, and Scholz E (2011) Who took the burden to answer on the meaning of left 
and right? Response behaviour on an open-ended question. In Conference Proceedings 
of the 64th Annual Conference of WAPOR: Public Opinion and the Internet. \\
Available at: \\
http://wapor.unl.edu/wp-content/uploads/2011/09/Zuell\_Scholz.docx

\medskip

\noindent
Z\"ull C, and Scholz E (2015) Who is willing to answer open-ended 
questions on the meaning of left and right?, {\em Bulletin of Sociological 
Methodology}, vol: 127(1), 26--42.

\end{document}